\documentclass[final,3p,times,twocolumn]{elsarticle}
\usepackage{amssymb}
\usepackage{lineno}

\usepackage[breaklinks,colorlinks,urlcolor=blue,linkcolor=blue,citecolor=blue]{hyperref}
\usepackage{color}

\journal{Nucl.\ Instr.\ and Meth.\ A}

\begin{document}

\begin{frontmatter}



\title{Assembly and Test of the Gas Pixel Detector for X-ray Polarimetry}


\author[thu,keylab]{H.~Li}
\author[thu,keylab]{H.~Feng\corref{cor}}
\ead{hfeng@tsinghua.edu.cn}
\cortext[cor]{Corresponding author}
\author[rome]{F.~Muleri}
\author[pisa]{R.~Bellazzini}
\author[pisa]{M.~Minuti}
\author[rome]{P.~Soffitta}
\author[pisa]{A.~Brez}
\author[pisa]{G.~Spandre}
\author[pisa]{M.~Pinchera}
\author[pisa]{C.~Sgr\`{o}}
\author[pisa]{L.~Baldini}
\author[thu,keylab]{R.~She}
\author[rome]{E.~Costa}

\address[thu]{Department of Engineering Physics and Center for Astrophysics, Tsinghua University, Beijing 100084, China}
\address[keylab]{Key Laboratory of Particle \& Radiation Imaging (Tsinghua University), Ministry of Education, China}
\address[rome]{IAPS/INAF, Via Fosso del Cavaliere 100, 00133 Rome, Italy}
\address[pisa]{INFN-Pisa, Largo B. Pontecorvo 3, 56127 Pisa, Italy}

\begin{abstract}
The gas pixel detector (GPD) dedicated for photoelectric X-ray polarimetry is selected as the focal plane detector for the ESA medium-class mission concept X-ray Imaging and Polarimetry Explorer (XIPE). Here we show the design, assembly, and preliminary test results of a small GPD for the purpose of gas mixture optimization needed for the phase A study of XIPE.  The detector is assembled in house at Tsinghua University following a design by the INFN-Pisa group. The improved detector design results in a good uniformity for the electric field. Filled with pure dimethyl ether (DME) at 0.8 atm, the measured energy resolution is 18\% at 6 keV and inversely scales with the square root of the X-ray energy. The measured modulation factor is well consistent with that from simulation, up to $\sim$0.6 above 6 keV. The residual modulation is found to be $0.30\% \pm 0.15\%$ at 6 keV for the whole sensitive area, which can be translated into a systematic error of less than 1\% for polarization measurement at a confidence level of 99\%. The position resolution of the detector is about 80 $\mu$m in FWHM, consistent with previous studies and sufficient for XIPE requirements. 
\end{abstract}


\end{frontmatter}

% \linenumbers

\section{Introduction}
\label{intro}

X-ray polarimetry is expected to be a powerful tool for astrophysics, offering extra information in addition to X-ray imaging, spectroscopy and timing.  It allows us to probe the magnetic field via synchrotron radiation or test the geometry via scattering, and is capable of testing fundamental physics such as quantum electrodynamics and general relativity under extreme magnetism or gravity \cite{Kallman2004,Bellazzini2010,Soffitta2013_xipe}. Despite the high demand in astrophysics, X-ray polarimetry has been an unexplored area for 40 years since the experiments on the OSO-8 satellite \cite{Weisskopf1976} in 1970s,  due to the absence of sensitive technology.  

Along with the development of micro-pattern gas detectors, it has become possible to image the tracks for electrons of a few keV in gas chambers, allowing for sensitive X-ray polarimetry depending on the photoelectric effect using the gas electron multiplier (GEM) with pixel readout \cite{Costa2001,Bellazzini2006_gpd}.  Compared with other readout techniques, the gas pixel detector (GPD) offers symmetric measurement in the two dimensions and delivers a low systematic error below one per cent even without instrument spinning \cite{Bellazzini2006_gpd,Muleri2012,Bellazzini2013}.  One of the key elements to the success of the GPD detector is the large-format high-resolution pixel readout chip. Several generations of dedicated ASIC chips have been developed, reaching to a pixel size of 50 $\mu$m and a chip size of 1.5cm $\times$ 1.5cm (105k pixels)  \cite{Bellazzini2004,Bellazzini2006_asic}.  Two versions of sealed test chambers, a small one \cite{Bellazzini2007_seal} and a large one \cite{Muleri2012}, were designed by the INFN-Pisa group and assembled by Oxford Instrument Analytical Oy.  The major improvement for the large version is that the background induced by the wall of the chamber is reduced and the electric field is more uniform \cite{Soffitta2012}. 

In 2015, the X-ray Imaging and Polarimetry Explorer (XIPE) \cite{Soffitta2013_xipe} was approved for phase A study by the European Space Agency (ESA). It is a space telescope concept dedicated to X-ray polarimetry in response to the call for medium-class missions. One of our tasks is to build refillable sealed GPD detectors to test and optimize the gas mixture of the detector. We therefore describe here the assembly of the GPD 
detectors and some preliminary test results. The optimization of the gas mixture and related tests of the detector will be reported in follow-up papers. Our work is based on the small version GPD, because it is relatively simple in structure and assembly, and sufficient for testing the gas mixture. 

\section{Detector structure and assembly}

\begin{figure}[t]
\centering
\includegraphics[width=\columnwidth]{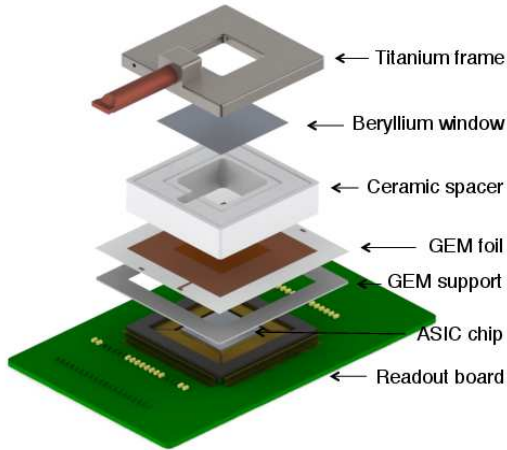}
\caption{A schematic drawing of the GPD. 
\label{fig:gpd}}
\end{figure}

\begin{figure}[t]
\centering
\includegraphics[width=\columnwidth]{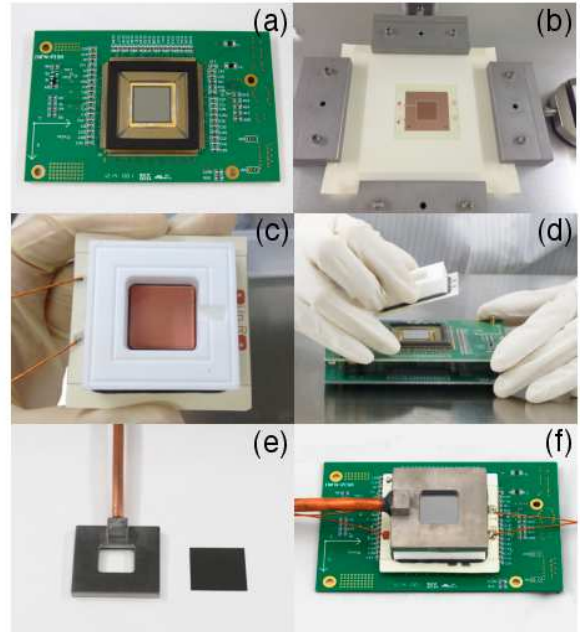}
\caption{Detector assembly. (a) ASIC chip in the ceramic package and both on the PCB; (b) GEM framing; (c) spacer and GEM; (d) gluing the spacer/GEM to the package; (e) titanium cap and beryllium window; (f) a completely assembled GPD.
\label{fig:assembly}}
\end{figure}

We start with the design of the small GPD \cite{Bellazzini2006_gpd,Bellazzini2007_seal}, see Figure~\ref{fig:gpd} for a schematic drawing. The ASIC chip \cite{Bellazzini2006_asic} is mounted using low outgassing silver epoxy inside a vacuum tight 304-pin ceramic package, which is soldered onto a printed circuit board (PCB). The bottom of the chip is connected to the electronic background and the heat from the chip is dissipated into the base of the package and then to the PCB board, where a Peltier cooler can be mounted on the other side and multiple via holes are used to increase the thermal conductivity.  The surface of the chip is measured to have an orientation tolerance of less than 20 $\mu$m with respect to the shoulder of the package using a measuring microscope. The distance between the chip surface and the package shoulder is 0.8 mm, which is the induction distance between the anode and the bottom layer of the GEM. The GEM foil is manufactured by SciEnergy Inc. It is 50 $\mu$m thick insulated by liquid crystal polymer; the laser-etched holes have a diameter of 30 $\mu$m and a pitch of 50 $\mu$m in a hexagonal pattern.  Four knife-edge clamps are used to apply tensions to the GEM foil and a FR4 frame is fixed on it to keep the tension using double sided sticky tape. The framed GEM foil is then placed right above the shoulder of the ceramic package; the frame is just outside the package and is in the air side so that the choice of its material is not important. A ceramic spacer of 1 cm thick (the drift distance) stands above the GEM foil and supports the cap of the chamber, which is a titanium plate with a 17mm $\times$ 17mm square hole sealed with a 100 $\mu$m thick beryllium window, both serving as the cathode of the detector.  On the top of the titanium plate, a copper tube with a diameter of 6 mm is mounted for vacuum pumping and gas filling. All the parts for the main body of the chamber are glued together using low outgassing epoxy and cured at a temperature of 60~$^\circ$C.  The copper tube is welded onto the titanium plate and some epoxy is used to further seal the joint.  After the chamber is filled with gas, the tube is cut and sealed using ultrasonic welding.  To change the gas mixture of the chamber, we will just cut the copper tube. Before filling the gas, the detector is pumped and baked for weeks until a vacuum down to $\sim$$10^{-9}$  mbar.  Some of the pictures during the assembly process and the final detector are shown in Figure~\ref{fig:assembly}. 

For all the tests in this paper, we adopt a gas mixture of pure dimethyl ether (DME) at 0.8 atm, for a direct comparison with previous results \cite{Muleri2010_performance}. The DME is purified to better than 0.99999 measured by gas chromatography.

%H20+O2 < 6ppm 
%CH4 < 0.5ppm
%CO2 < 1.5 ppm
%N2 < 2 ppm

\section{Uniformity of the electric field}

\begin{figure}[t]
\centering
\includegraphics[width=0.49\columnwidth]{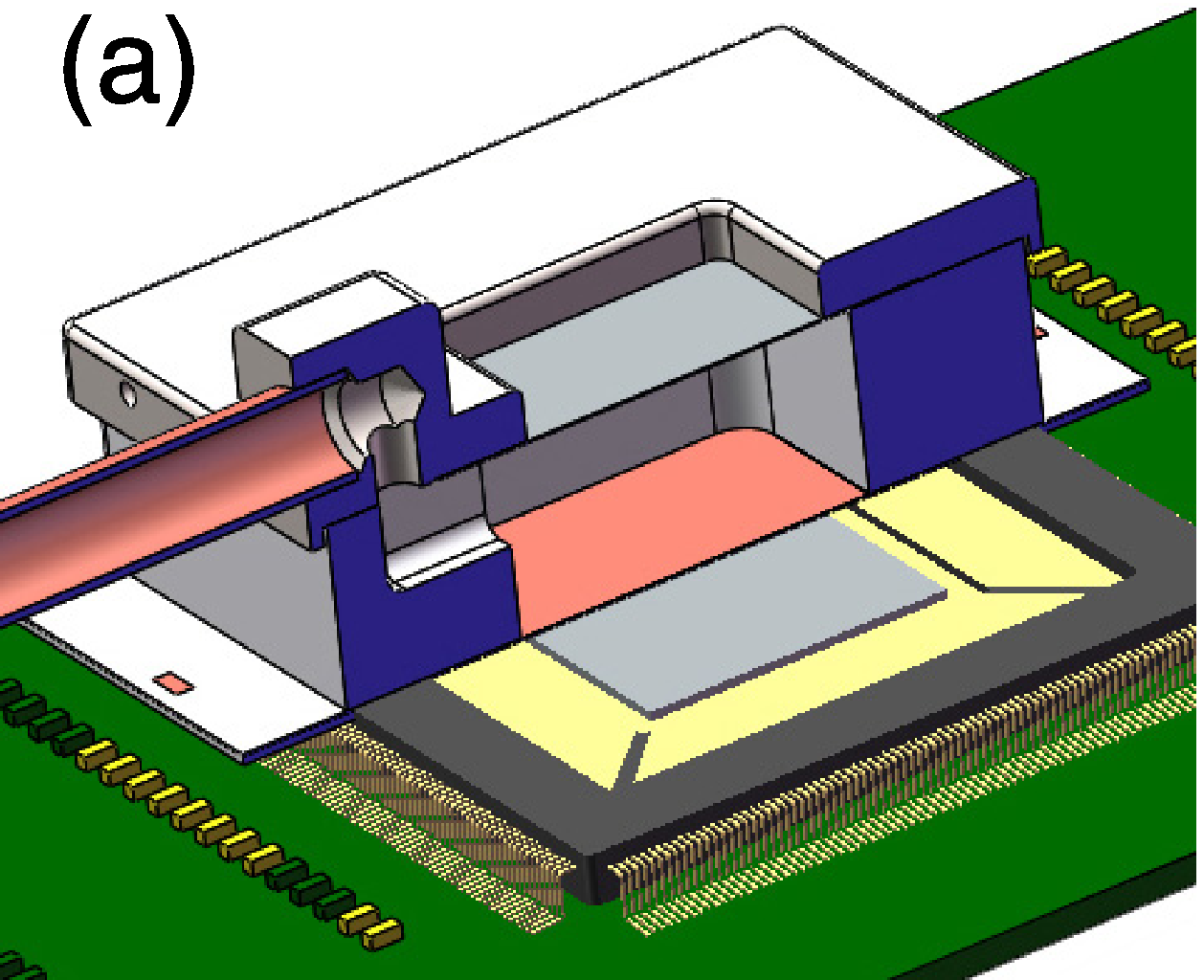}
\includegraphics[width=0.49\columnwidth]{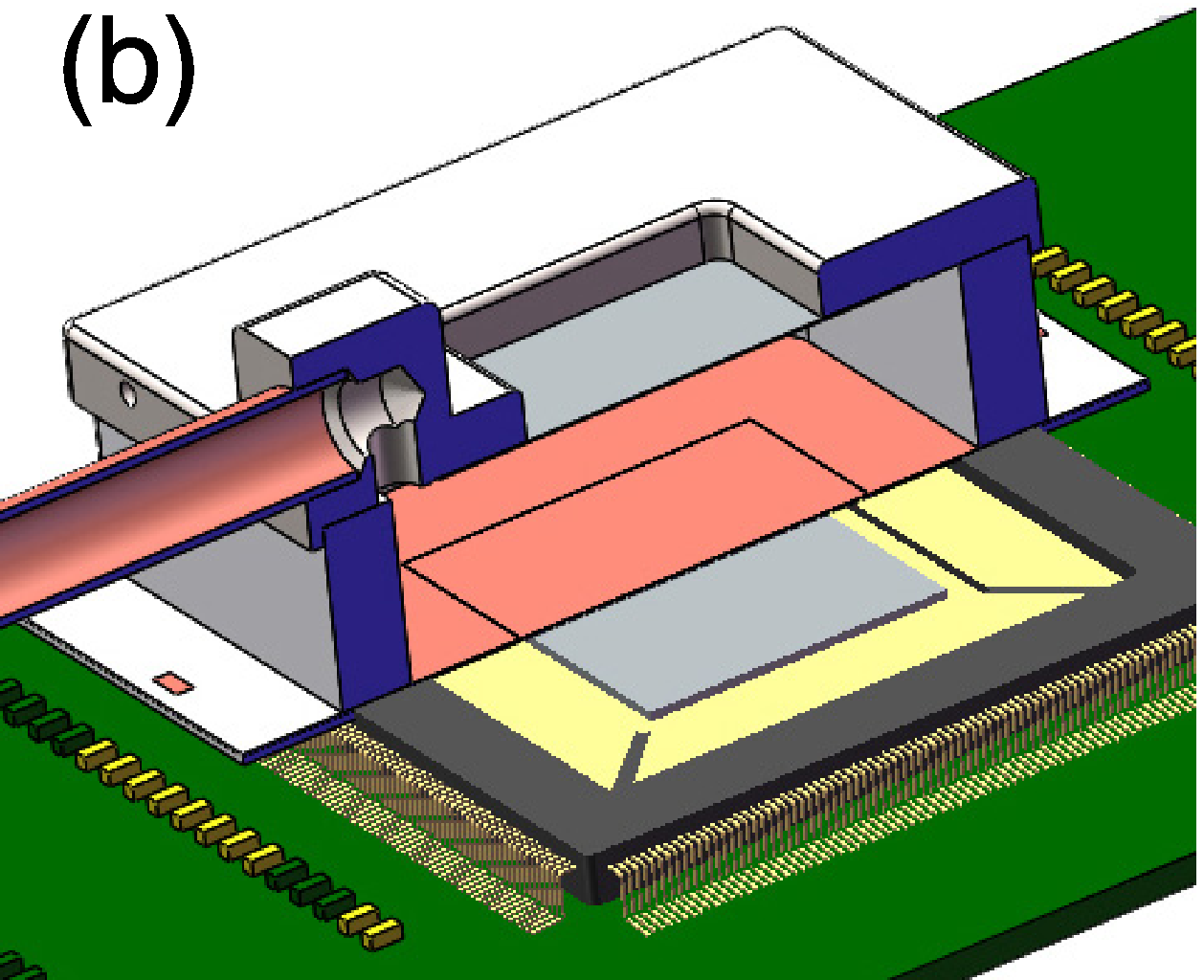}
\caption{Inner structure of the chamber that connects the copper tube and the inner chamber, for the original design (left) and the improved design (right). 
\label{fig:pipe}}
\end{figure}

\begin{figure}[t]
\centering
\includegraphics[width=\columnwidth]{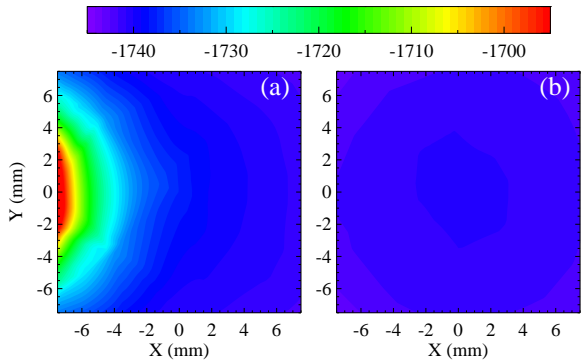}
\caption{Intensity map of the electric potential in the middle plane of the chamber, for the original design (left; Figure~\ref{fig:pipe}a) and the improved design (right; Figure~\ref{fig:pipe}b). 
\label{fig:maxwell}}
\end{figure}

\begin{figure}[!h]
\centering
  \includegraphics[width=0.9\columnwidth]{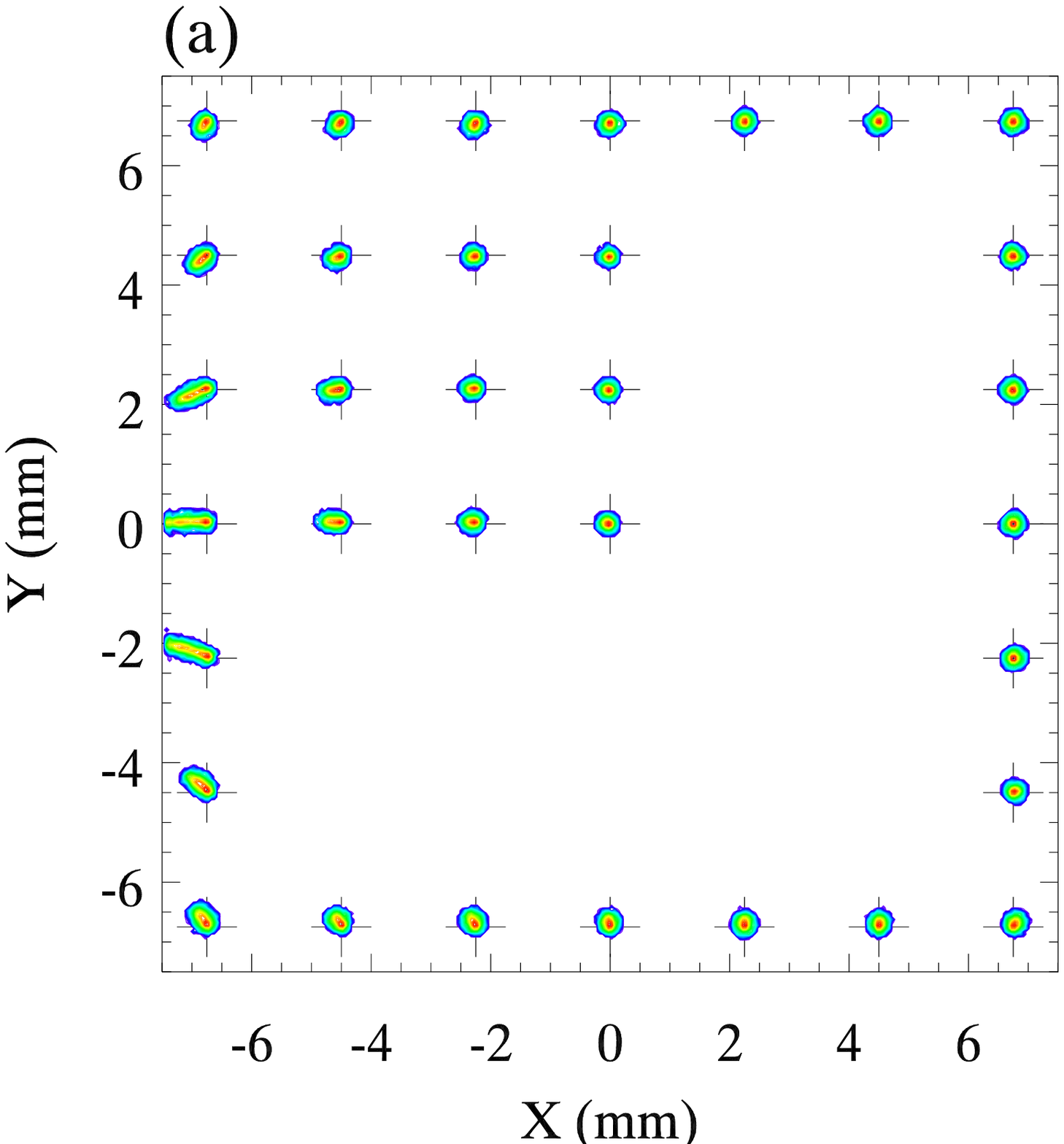}\\
  \includegraphics[width=0.9\columnwidth]{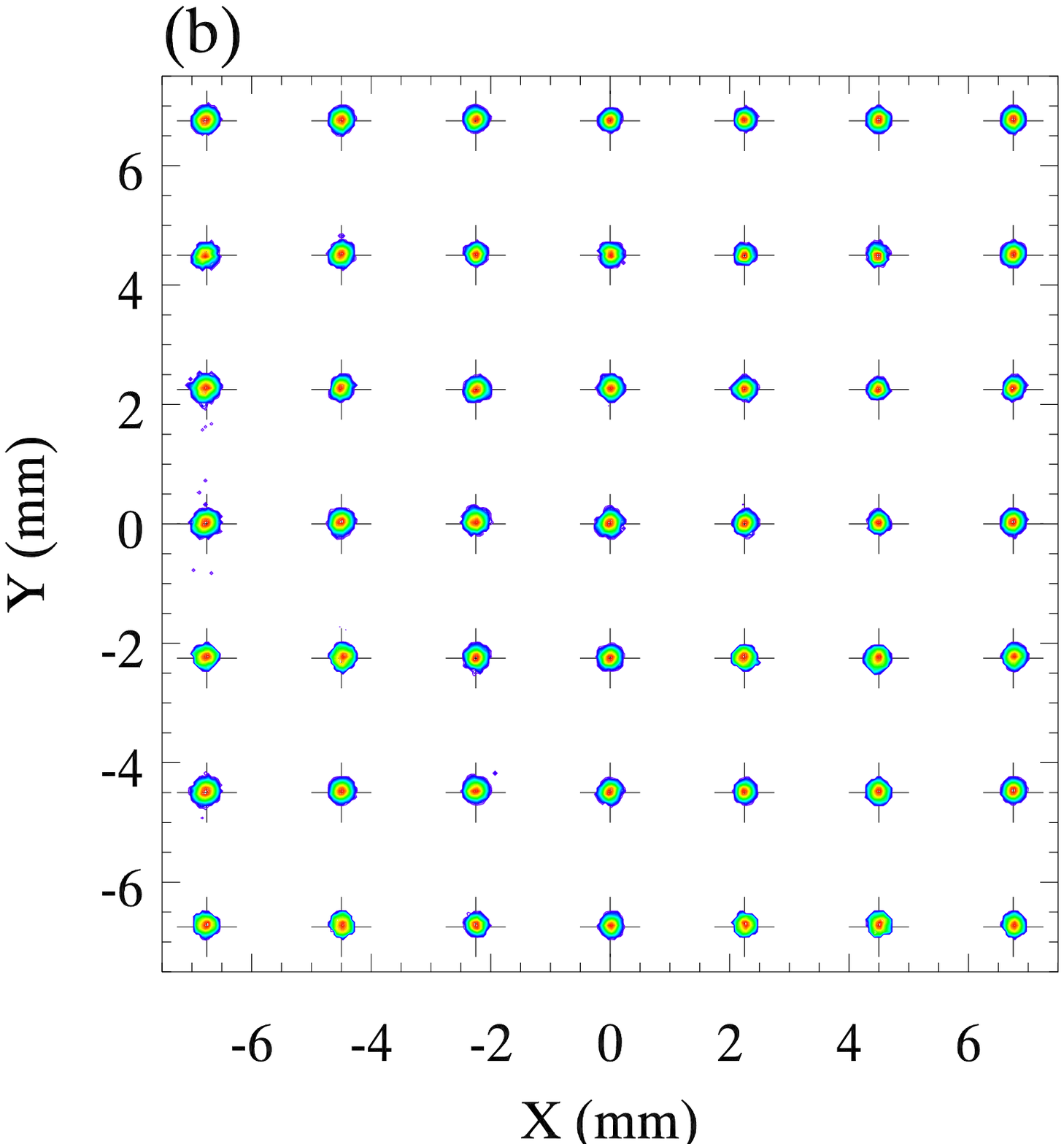}
\caption{Surface uniformity test using a narrow beam for the original design (top; Figure~\ref{fig:pipe}a) and the improved design (bottom; Figure~\ref{fig:pipe}b). The crosses indicate the incident positions for each beam. The contours are the collected charge distribution for each beam test, shown in log scale.}
\label{fig:surface}
\end{figure}

For the small GPD design, a major challenge is to improve the uniformity of the electric field, which will directly influence the electron drift and consequently the accuracy of the electron track measurement. In our first design (Figure~\ref{fig:pipe}a), a slot on the thick wall of the spacer and a right-angle hole on the titanium cap is machined to connect the copper tube and the inner chamber. However, such a design creates an asymmetric structure and leads to large non-uniformity for the electric field. Figure~\ref{fig:maxwell}a shows the intensity map of the electric potential in the middle plane along the drift direction ($Z = 5$~mm above the GEM top plane) in the GPD chamber simulated using Ansoft Maxwell, assuming a voltage of -2740 V on the drift plane and -740 V on the top plane of the GEM. As one can see, the electric field is quite uneven toward the tube side.  To test the uniformity, we constructed a narrow beam of X-rays and illuminated the detector at different locations. At positions close to the tube side, the cumulative charge map produced by all the events is elongated, indicative of a distorted electric field (Figure~\ref{fig:surface}a). The test did not cover the whole sensitive area because the non-uniformity is already obvious and a full coverage takes a huge amount of time. 

To correct the field distortion, we changed the design of the interconnection structure. The spacer wall becomes thinner and stands outside of the right-angle hole in the titanium plate (Figure~\ref{fig:pipe}b). This makes the ceramic spacer look the same on each face from inside. The simulated potential map shows that the maximum differential voltage in the middle plane is only 2.4 volt or 0.14\% (Figure~\ref{fig:maxwell}b).  The narrow beam test indicates that the charge distribution is no longer elongated or position sensitive (Figure~\ref{fig:surface}a). The uniformity of the electric field is much improved and such a design is capable of testing the performance of the detector over the whole sensitive area. 

\section{Test and results}

In order to characterize the detector performance, we constructed a desktop X-ray source including an X-ray tube manufactured by Oxford Instrument and crystals to produce fully polarized monochromatic X-ray beams by Bragg diffraction at  45 degrees, following the setup described by Muleri et al.\ (2008,2010) \cite{Muleri2008_bragg,Muleri2010_performance}. The choice of crystals and their Bragg diffracted energies at 45 degrees are listed in Table~\ref{tab:bragg}.  The bremsstrahlung continuum from the X-ray tube serves as the incident beam for diffraction. A capillary plate manufactured by Hamamatsu, which has a hole diameter to thickness ratio of 1/167, is mounted in the light path after the diffraction to select photons due to diffraction rather than Compton scattering. With such a source, we are able to produce polarized X-rays at 2.67, 3.74, 5.33, 6.09, and 7.49~keV. 

\begin{table}[t]
  \centering
  \caption{The choice of Bragg crystals and diffraction orders, corresponding X-ray energies, and measured modulation factors.}
  \label{tab:bragg}
  \begin{tabular}{cccc}
    \noalign{\smallskip}\hline\noalign{\smallskip}
    $E$ (keV) & crystal & order & $\mu$ (\%)\\
    \noalign{\smallskip}\hline\noalign{\smallskip}
    2.67 & MgF$_2$ &1st & 27.2$\pm$1.3 \\
    3.74 & Al & 1st & 42.3$\pm$0.4 \\
    5.33 & MgF$_2$ & 2nd &54.1$\pm$0.6 \\
    6.09 & LiF & 2nd & 58.2$\pm$0.7 \\
    7.49 & Al & 2nd & 61.9$\pm$1.3 \\
    \noalign{\smallskip}\hline\noalign{\smallskip}
  \end{tabular}
\end{table}

\begin{figure*}[bt]
\centering
\includegraphics[width=0.32\textwidth]{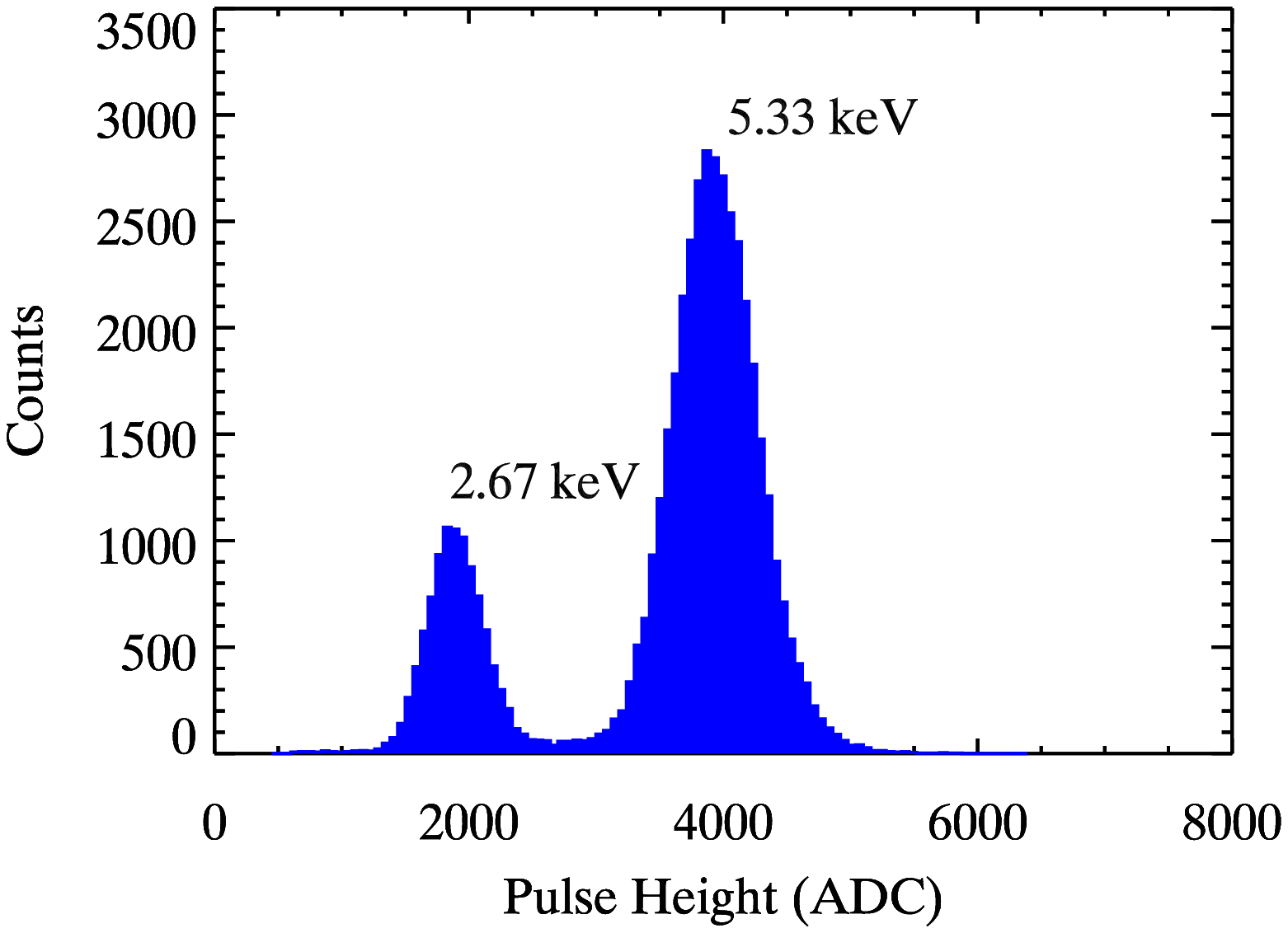}
\includegraphics[width=0.32\textwidth]{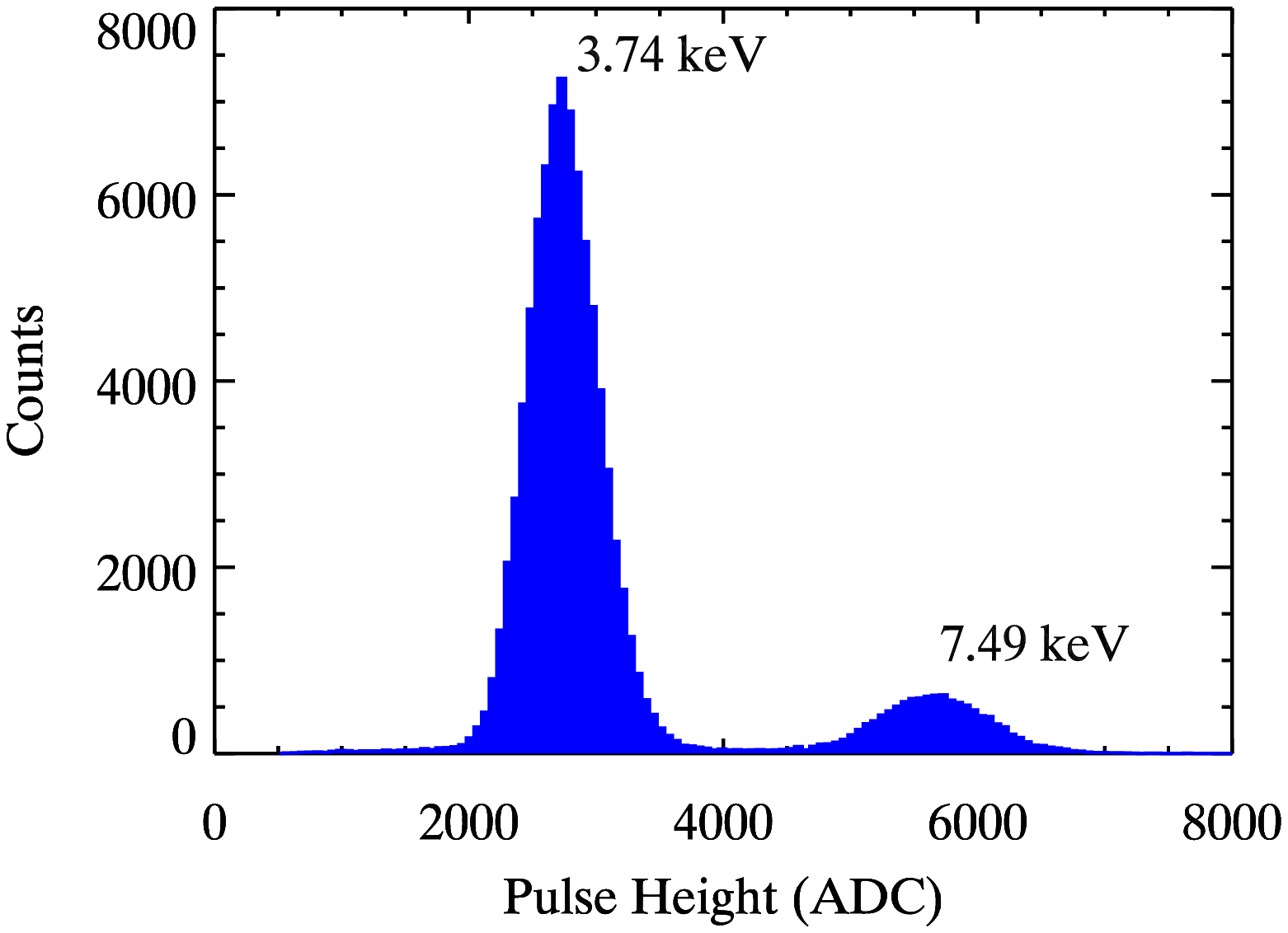}
\includegraphics[width=0.32\textwidth]{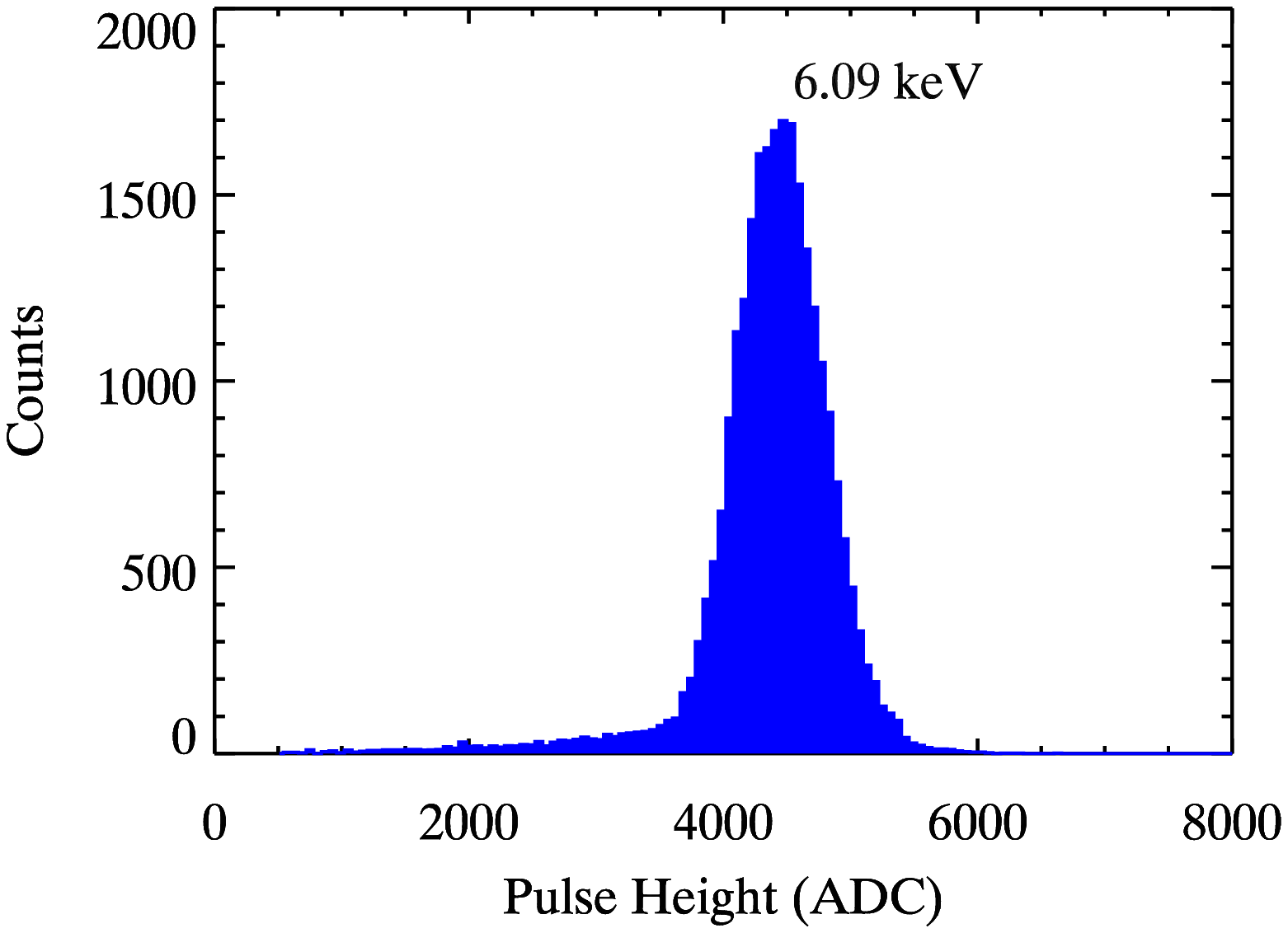}
\caption{Pulse height distribution measured with X-rays created by 45-degree Bragg diffraction with MgF$_2$ (left), Al (middle) and LiF (right) crystals. 
\label{fig:spec}}
\end{figure*}

\begin{table}[t]
  \centering
  \caption{Spectral fitting results at different energies.}
  \label{tab:energy}
  \begin{tabular}{cccc}
    \noalign{\smallskip}\hline\noalign{\smallskip}
    $E$ & Peak & FWHM & $\Delta E / E$ \\
    (keV) & (ADC) & (ADC) & (\%) \\
    \noalign{\smallskip}\hline\noalign{\smallskip}
    2.67 & $1901.1 \pm 2.3$  & $230.0 \pm 1.8$ & $26.10 \pm 0.23$ \\
    3.74 & $2744.9 \pm 0.9$  & $278.2 \pm 0.7$ & $22.44 \pm 0.06$ \\
    5.33 & $3938.7 \pm 1.8$  & $331.7 \pm 1.4$ & $18.97 \pm 0.08$ \\
    6.09 & $4548.6 \pm 2.3$  & $358.1 \pm 1.9$ & $17.83 \pm 0.10$ \\
    7.49 & $5650.3 \pm 5.4$  & $402.0 \pm 4.9$ & $16.23 \pm 0.21$ \\
    \noalign{\smallskip}\hline\noalign{\smallskip}
\end{tabular}
\end{table}

\subsection{Spectral response}

The incident X-rays have a narrow beam size due to collimation with the capillary plate.  The beam is aligned to illuminate the central region of the detector. The analysis of the data is implemented in the PIXY package developed by the INFN-Pisa group. We select events in a circular aperture with a radius of 1 mm.  A noise threshold of 8 ADC is adopted. The energy for each event is calculated by summing all of the pixel values after the noise cut. The energy spectra are shown in Figure~\ref{fig:spec}. The photopeaks due to the 1st and 2nd orders of diffraction with the MgF$_2$ and Al crystals are well distinguished.  The energy spectra also show a tiny low energy tail (e.g., see the 6.09 keV spectrum), which is more prominent for high energy photons and is caused by absorption of the photoelectron on the GEM foil or the beryllium window that terminates further ionization in the gas \cite{Muleri2012}. To extract the spectral information, we fit a Gaussian function to each peak with results listed in Table~\ref{tab:energy}.  

\begin{figure}[t]
\centering
\includegraphics[width=\columnwidth]{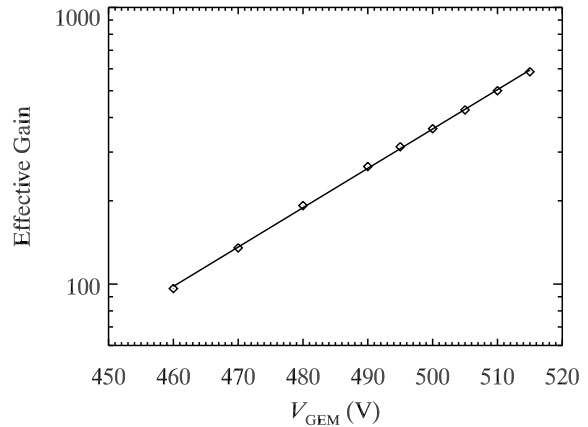}
\caption{Effective gain of the GEM versus the GEM voltage. 
\label{fig:gain}}
\end{figure}

\begin{figure}[t]
\centering
\includegraphics[width=\columnwidth]{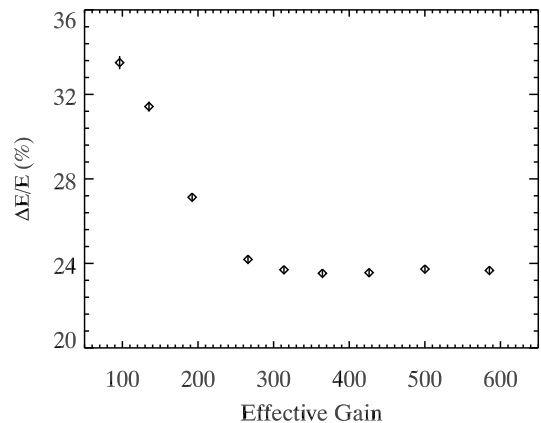}
\caption{Fractional spectral resolution (FWHM/$E$) at 3.74 keV versus the effective gain of the GEM.
\label{fig:fwhm_gain}}
\end{figure}

The drift field is fixed at 2 kV/cm, where the coefficient of transversal diffusion for electron drifting is minimized in pure DME at 0.8 atm.  The induction field (or the collection field) is set to 3 kV/cm; a smaller field may lower the fraction of electrons collected by the anode, and a higher field may lead to multiplication in the induction region.  Figure~\ref{fig:gain} shows the change of the effective gain as a function of the GEM voltage, and Figure~\ref{fig:fwhm_gain} illustrates the variation of the spectral resolution along with the change of the effective gain.  We choose a GEM voltage of 500 V where the effective gain is near 400, and the spectral resolution is minimum and insensitive to the change of the gain. We note that the choice of this GEM voltage is also optimal for polarimetry, resulting in a maximum degree of modulation, which is not shown here. 

\begin{figure}[t]
\centering
\includegraphics[width=0.43\textwidth]{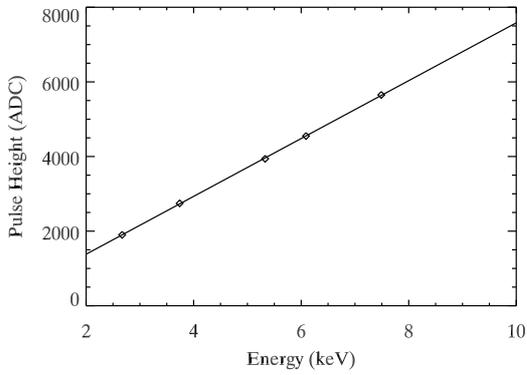}
\caption{The pulse hight of the photopeak versus the incident X-ray energy.
\label{fig:adcvsen}}
\end{figure}

\begin{figure}[t]
\centering
\includegraphics[width=0.43\textwidth]{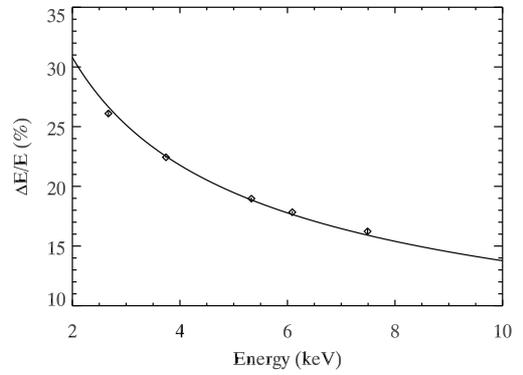}
\caption{Measured energy resolution ($\Delta E$ = FWHM) versus the X-ray energy, along with a best-fit curve in the form of $\Delta E/E \propto 1/\sqrt{E}$. 
\label{fig:fwhm}}
\end{figure}

The detector exhibits a good linearity between the incident X-ray energy and the output pulse height (see Figure~\ref{fig:adcvsen}).  The measured energy resolution (Figure~\ref{fig:fwhm}) in terms of FWHM/$E$ is found to be scaled with the square root of the X-ray energy, consistent with that the noise is dominated by Fano fluctuation. 

\subsection{Polarimetry}

\begin{figure}[t]
\centering
\includegraphics[width=0.49\columnwidth]{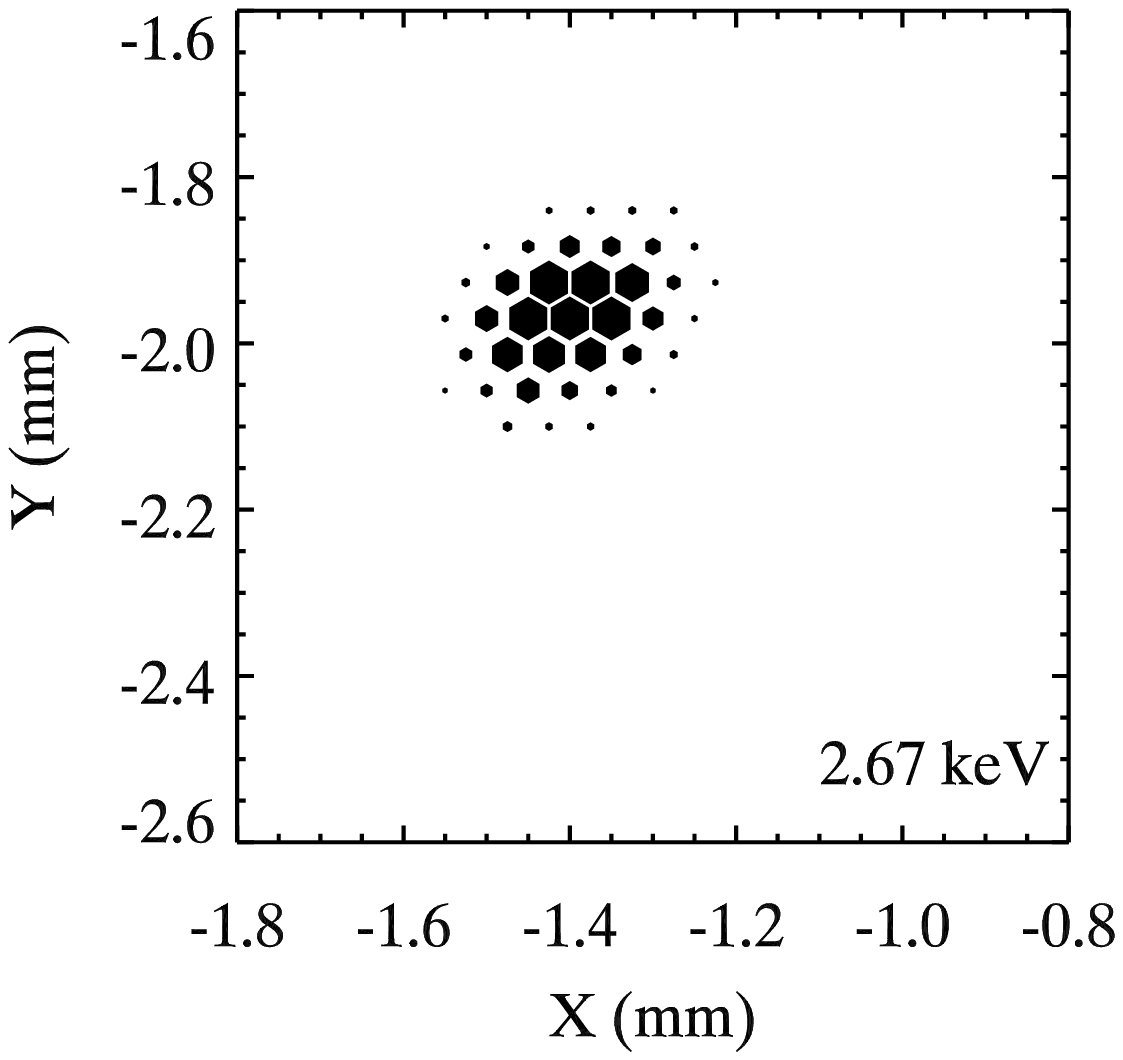}\\
\includegraphics[width=0.49\columnwidth]{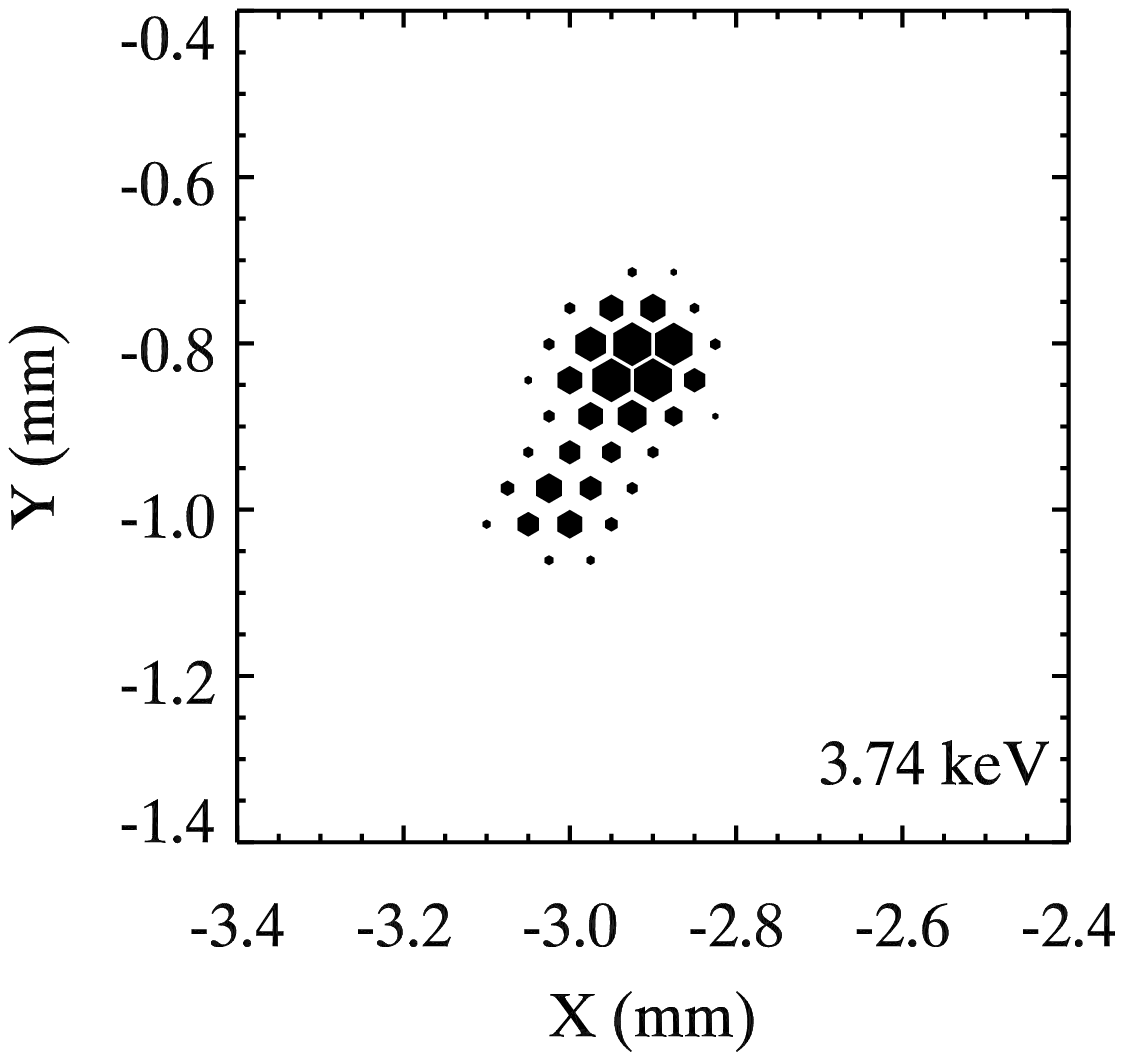}
\includegraphics[width=0.49\columnwidth]{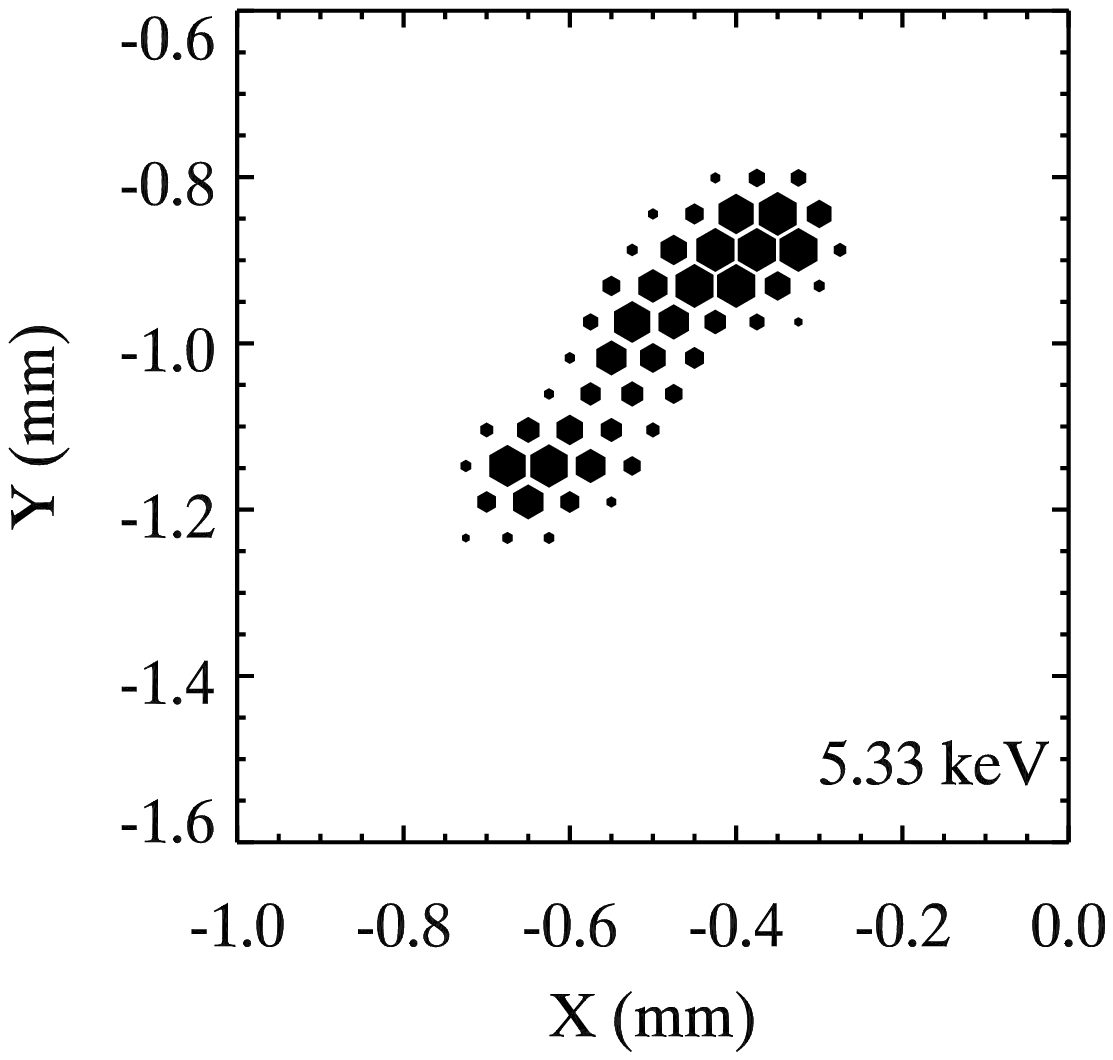}\\
\includegraphics[width=0.49\columnwidth]{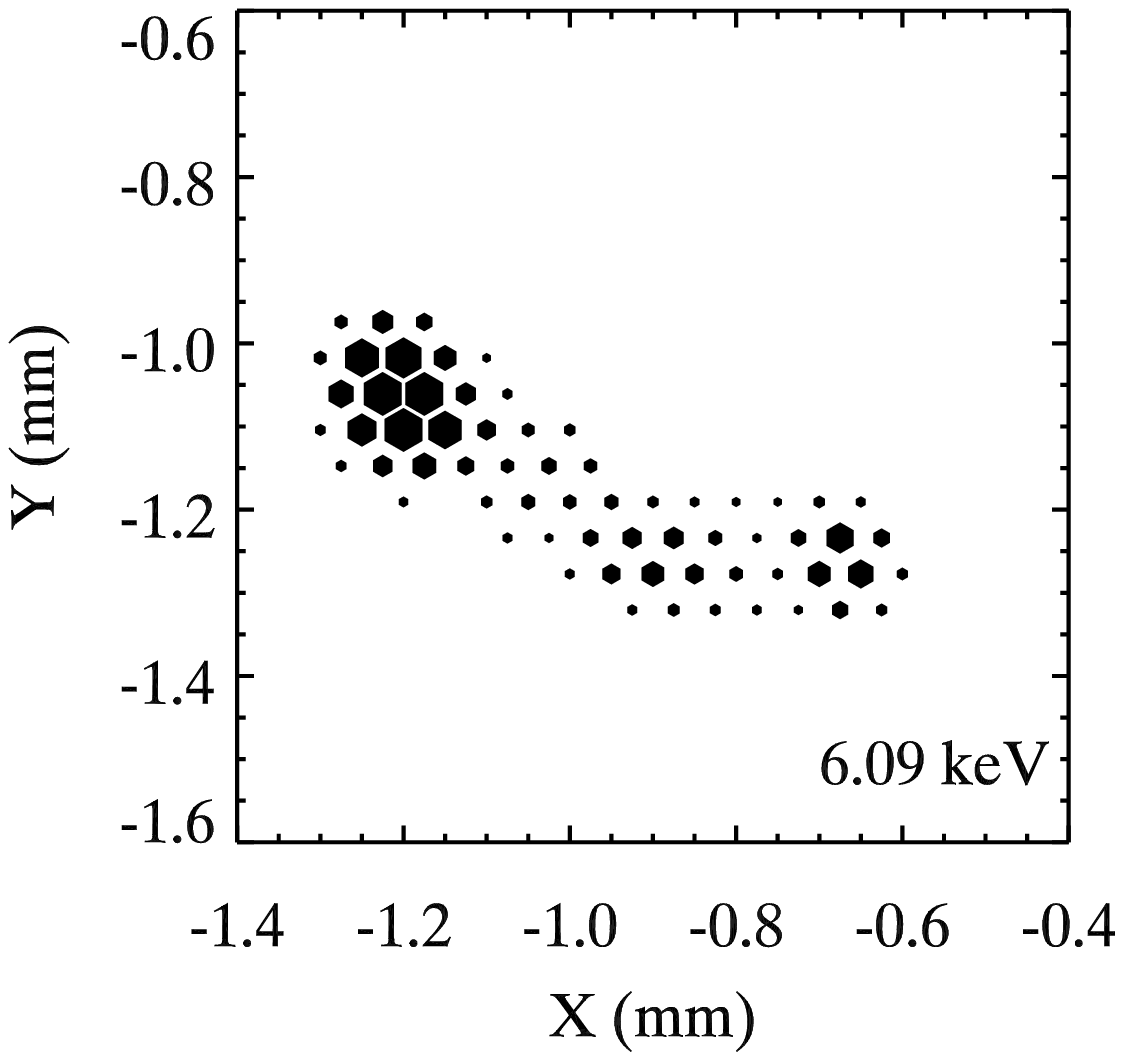}
\includegraphics[width=0.49\columnwidth]{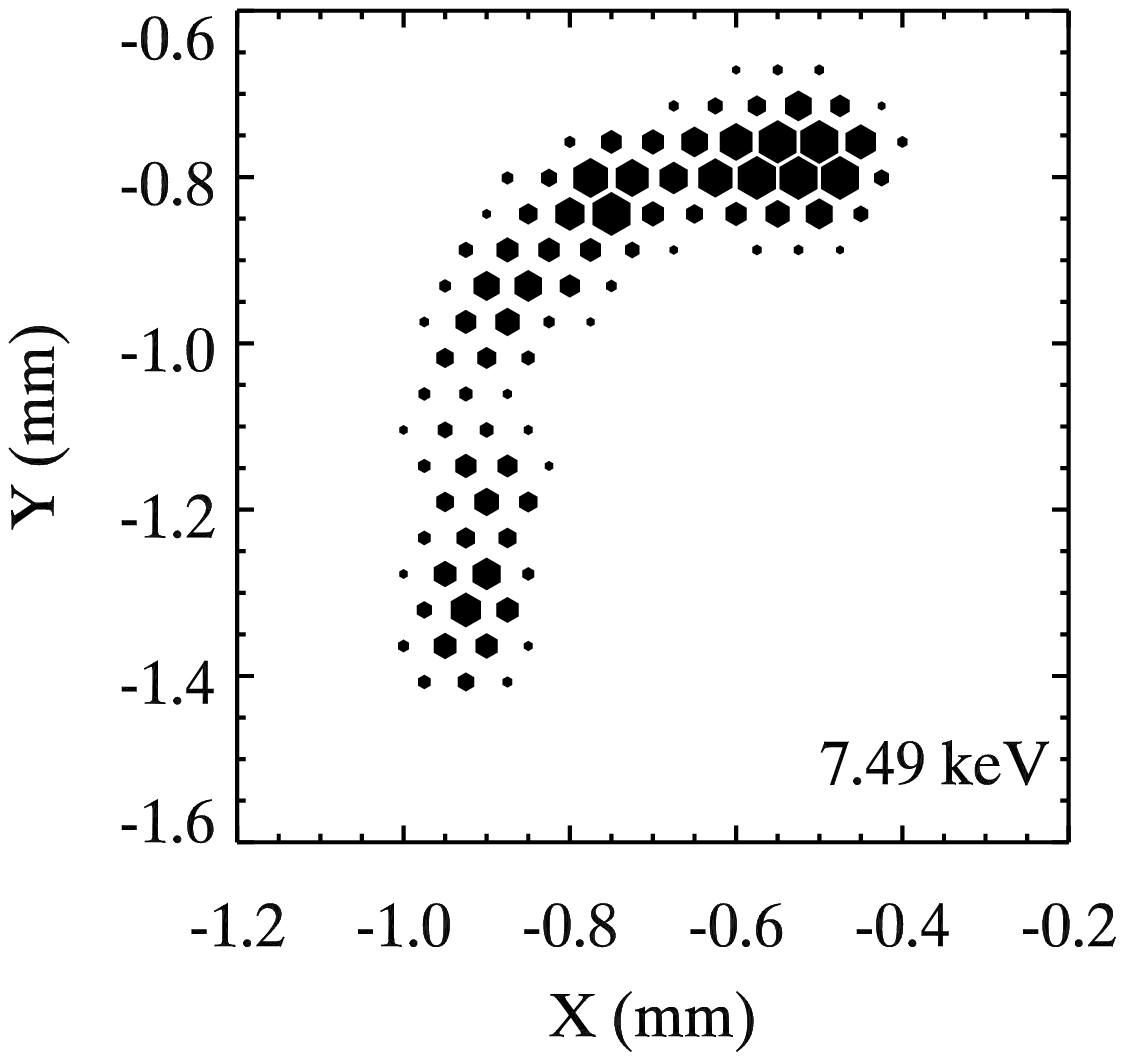}
\caption{Example electron tracks produced by X-rays of different energies.
\label{fig:track}}
\end{figure}

Examples of electron tracks produced by X-rays of different energies are displayed in Figure~\ref{fig:track}.  For polarimetric analysis, events with two or more separate clusters of charges are not used. Those having a total hit of less than 27 pixels are removed too, as they may cause a systematic effect at $\pm$60 degrees due to the pattern of the GEM holes.   For a direct comparison with previous results, we further discard 25\% of events with low eccentricity following Muleri et al.\ (2010)  \cite{Muleri2010_performance}.  The emission angle of photoelectrons is derived by finding the principal axis of the whole track for events with energies below 3 keV, or using the so-called impact point method \cite{Bellazzini2003_sim} for high energy events. 

\begin{figure}[t]
\centering
\includegraphics[width=0.8\columnwidth]{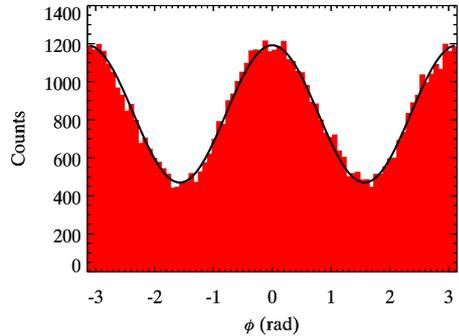}
\caption{Modulation curve measured with a fully polarized beam of 3.74~keV. The solid line is the best-fit cosine function with a degree of modulation of $42.3\% \pm 0.4\%$.
\label{fig:modcurve}}
\end{figure}

\begin{figure}[t]
\centering
\includegraphics[width=\columnwidth]{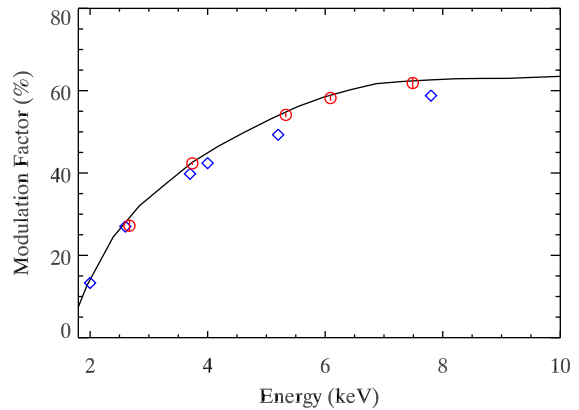}
\caption{Modulation factor versus energy. The red circles indicate results measured with our detector, while the blue diamonds are results obtained by a previous detector with the same gas mixture \cite{Muleri2010_performance}.  The solid curve is the result obtained using GEANT4/Garfield simulations. 
\label{fig:mf}}
\end{figure}

A modulation curve, which is the histogram of photoelectron emission angles, measured with a fully polarized beam of 3.74 keV is shown in Figure~\ref{fig:modcurve} . The degree of modulation is defined as the fractional peak-peak amplitude of the modulation curve, $(N_{\rm max} - N_{\rm min}) / (N_{\rm max} + N_{\rm min})$. The modulation factor $\mu$ is the degree of modulation in response to a fully polarized beam and is one of the most important parameters for a polarimeter, which is directly scaled with the sensitivity for polarimetry.  The measured modulation factors at all of the available energies are listed in Table~\ref{tab:bragg} and displayed in Figure~\ref{fig:mf}. For comparison, the results obtained by Muleri et al.\ (2010) with the same gas mixture are also shown \cite{Muleri2010_performance}. The curve in the figure represents the expected modulation factor versus energy calculated using GEANT4/Garfield simulations taking into account the transversal electron diffusion during drift, multiplication and collection, electronics noise, and the resolution of the readout chip.  Our result is well consistent with that obtained with the simulation. 

\begin{figure}[!h]
\centering
\includegraphics[width=0.8\columnwidth]{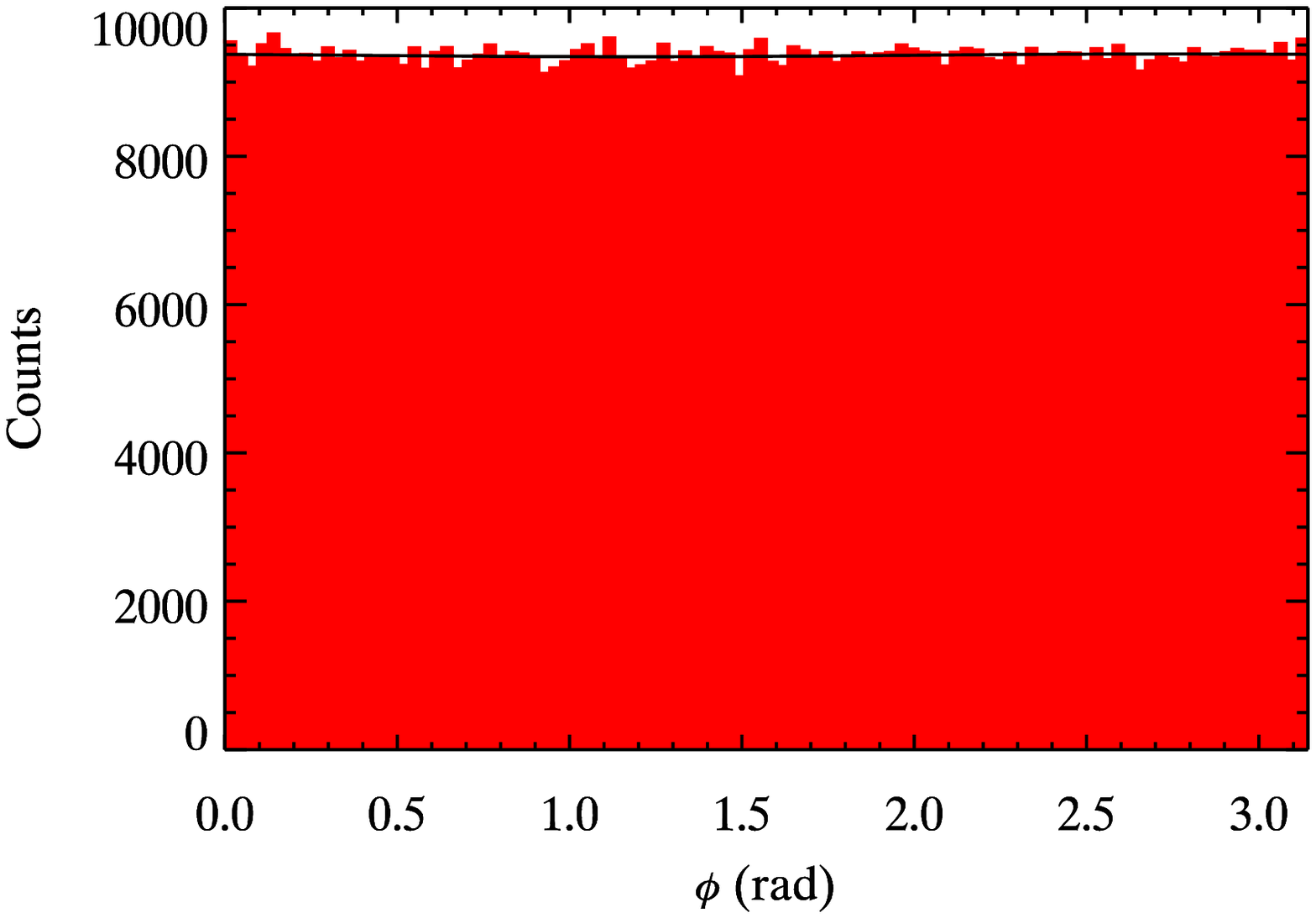}\\
\includegraphics[width=0.8\columnwidth]{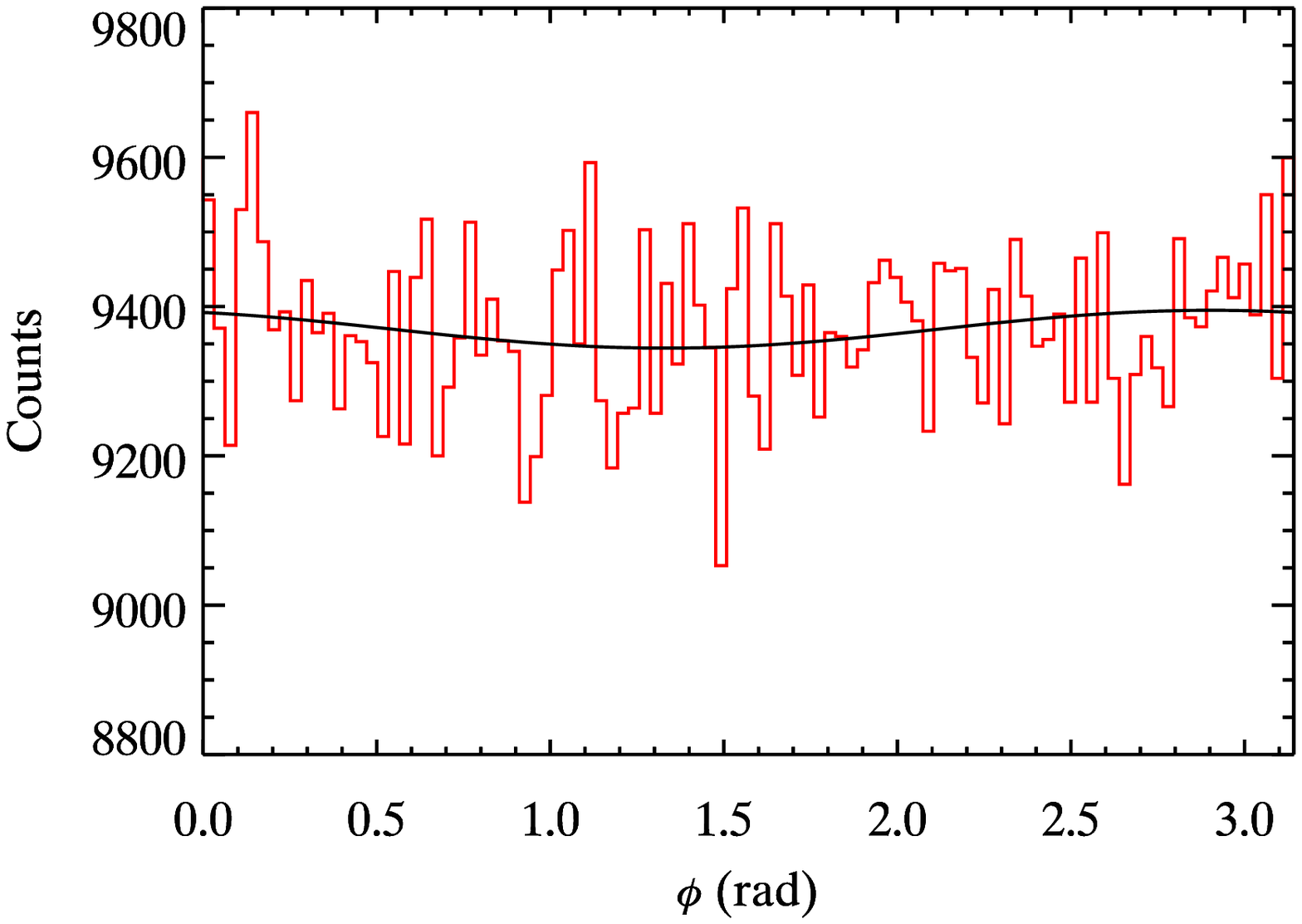}
\caption{Residual modulation measured with an unpolarized beam at 5.9~keV. The degree of modulation is $0.30\% \pm 0.15\%$, corresponding to a 99\% upper limit of 0.55\% in degree of modulation or 0.94\% in degree of polarization. The bottom panel is a zoom-in of the top panel.
\label{fig:unpol}}
\end{figure}

To test the systematics of the detector, we constructed an unpolarized source by illuminating a manganese target using the same X-ray tube, which produces fluorescent lines at 5.9 keV. To further smooth out any possible polarized component in the X-ray beam, e.g., due to scattering, we kept rotating the source slowly around its beam axis during the test by 1 degree per second for some 80 complete cycles.  About 960,000 valid events were detected by the detector, almost evenly distributed on the sensitive area, 15mm $\times$ 15mm. This, given a modulation factor of 0.58 at 5.9 keV, corresponds to a minimum detectable polarization (MDP) of 0.75\% at 99\% confidence level. Figure~\ref{fig:unpol} shows the measured modulation curve for the unpolarized beam. The measured degree of modulation is $0.30\% \pm 0.15\%$, corresponding to a 99\% upper limit of 0.55\% in degree of modulation or 0.94\% in degree of polarization by taking into account the modulation factor.  The result is consistent with statistical fluctuation from zero polarization, and indicates that the instrument systematic error is below 1\% for polarimetric measurement.

\subsection{Imaging}

\begin{figure}[t]
\centering
\includegraphics[width=\columnwidth]{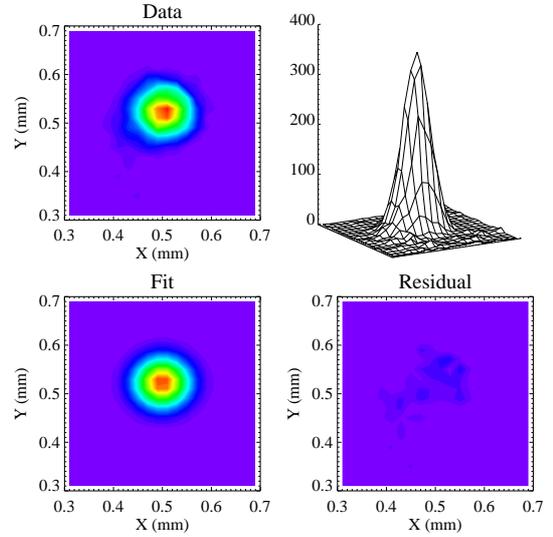}
\caption{Image of a narrow X-ray beam detected by the GPD and fitting with a 2D Gaussian profile. The beam size is around 30 $\mu$m in FWHM and the intrinsic point spread function is measured to have a FWHM of about 78 $\mu$m.
\label{fig:spot1}}
\end{figure}

\begin{figure}[!h]
\centering
\includegraphics[width=0.8\columnwidth]{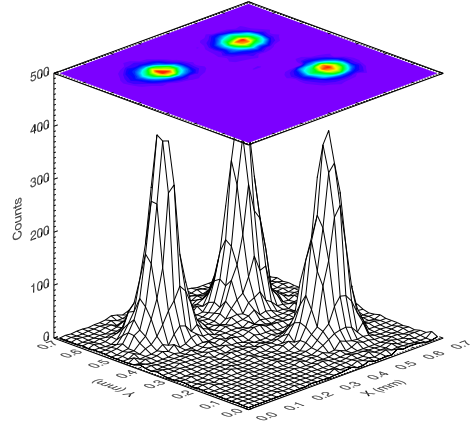}
\caption{Image of three sources with 300 $\mu$m apart in $X$ and $Y$.
\label{fig:spot3}}
\end{figure}

We constructed a narrow X-ray beam to characterize the imaging properties of the GPD. The X-ray tube has a focal spot size of about 100 $\mu$m. We use a stainless steel diaphragm with a diameter of 30 $\mu$m mounted at the end of a 18 cm long tube to collimate the source. The beam size is measured by scanning the beam with a Si-Pin detector at the same distance of the GPD, which is half masked with a sharp edge \cite{Soffitta2013_imaging}. Assuming that the intensity distribution of the X-ray beam can be described by a 2D Gaussian, an error function is adopted to fit the variation of the counting rate versus the scanning position. The beam size in FWHM is measured to be 31.8 $\mu$m in one direction and 28.6 $\mu$m in the orthogonal direction, consistent with the size of the diaphragm.

Figure~\ref{fig:spot1} shows the detected image by the GPD with the narrow beam. The X-ray interaction point is adopted for image reconstruction. A 2D Gaussian function is adequate to fit the data with $\chi ^2$ = 396.7 and 393 degrees of freedom. The measured FWHM of the point spread function is $87.4 \pm 0.9$ $\mu$m in the $X$ direction and $79.8 \pm 0.9$ $\mu$m in the $Y$ direction.  Taking into account the contribution of the beam size, the intrinsic spatial resolution is about 78 $\mu$m in FWHM.  We take three exposures by shifting the beam by 300 $\mu$m in two orthogonal directions.  Figure~\ref{fig:spot3} shows an image by combining all of the data, to simulate an image of three nearby point-like sources.

\section{Discussion and Conclusion}

In this paper, we presented the assembly of the GPD and some preliminary test results. We demonstrate that we are able to assemble detectors qualified for the phase A study of the XIPE mission. The improved design of the detector structure results in a good uniformity for the electric field.  Both the simulation with Maxwell and test using a narrow X-ray beam suggest that the electric field for the improved design is capable of X-ray polarimetry with low systematics. To compare with previous results, mainly with Muleri et al.\ (2010) \cite{Muleri2010_performance}, we choose a gas mixture of 0.8 atm pure DME that is relatively easy to fill (no need to mix with other gas components). The GEM foils used in the two studies are different. Our GEM is 50 $\mu$m thick with holes of 30 $\mu$m in diameter and 50 $\mu$m pitch, while their GEM is 100 $\mu$m thick with holes of 48 $\mu$m in diameter and 80 $\mu$m pitch.  The choice of drift field, GEM voltage, and the induction field is to maximize the imaging precision and signal to noise ratios. 

The spectral resolution of our detector at 6 keV is better than that reported in Muleri et al.\ (2010), 18\% versus 24\% \cite{Muleri2010_performance}, which may be a direct consequence of the impurity of the DME gas.  We once filled a detector and obtained worse resolution, but it was improved after we resolved the outgassing problem of the gas filling tubes between the gas cylinder and the detector.  They also obtained nice spectral resolutions, e.g., 15\% \cite{Bellazzini2007_seal} or 19\% \cite{Muleri2012} at 6 keV with a different gas mixture (DME:He 8:2 at 1 bar). Thus, the previous low resolution with pure DME is not due to the nature of the gas, rather it could be due to the non-uniformity of the GEM gain at different positions, as their beam size is much larger than ours. Once filled and sealed, our detector does not show a decline of performance with time, at least on a timescale of a year or so, which will be reported in detail later. 

The modulation factors that we measured is slightly higher than in Muleri et al.\ (2010) (Figure~\ref{fig:mf}). We note that the impurity of gas, which leads to electron attachment and loss during drifting, is not crucial to the modulation measurement. We found that consistent modulation factors could be obtained with the same detector filled with gases of different purity, e.g., with an energy resolution of 18\% to $\sim$30\%.  We note that the difference on the GEM thickness and hole resolution by the two studies, according to simulation, has almost no effect on the modulation factor with the choice of such a gas mixture.

The residual modulation, measured to be $0.30\% \pm 0.15\%$, is consistent with the value 0.54\% obtained by Muleri et al.\ (2012) \cite{Muleri2012} and $0.18\% \pm 0.14\%$ as reported by Bellazzini et al.\ (2013) \cite{Bellazzini2013}, although the two previous studies used a different gas mixture (DME:He 8:2 at 1 bar).  The measurement by Muleri et al.\ (2012) \cite{Muleri2012} should be dominated by statistical errors, because the number of counts we collected is much higher than what they got (960k versus 125k).  Limited by the source intensity, our measurement for the residual modulation covers the whole sensitive area of the detector. Otherwise, it would take too much time to get a meaningful measurement just in the focal point region using our existing setup. If we extract events in the central region of 3mm in radius, the residual modulation is $0.15\% \pm 0.39\%$ constructed from 129k counts, which is dominated by statistical errors. 

The imaging performance of our detector is well consistent with that measured by Soffitta et al.\ (2013) \cite{Soffitta2013_imaging}. A position resolution of about 80 $\mu$m in FWHM corresponds to an angular resolution of 4.7 arcsec given a telescope with a focal length of 3.5 m, sufficient to sample the point spread function for the XIPE optics which is required to have a angular resolution of less than 30 arcsec. 

To conclude, our assembled GPD shows a performance consistent with that of the best ones in previous studies, which will enable us to test and optimize the gas mixture for the phase A study of XIPE. 

\section*{Acknowledgements}

HF acknowledges funding support from the National Natural Science Foundation of China under grant No.\ 11222327, and the Tsinghua University Initiative Scientific Research Program.

%%Reference
%\bibliographystyle{model1a-num-names} 
%\bibliography{gpd}

\end{document}